\documentclass[aps,prb,twocolumn,10pt,superscriptaddress,showpacs]{revtex4-1}

\def\bra#1{\mathinner{\langle{#1}|}}
\def\ket#1{\mathinner{|{#1}\rangle}}
\def\braket#1{\mathinner{\langle{#1}\rangle}}

\let\protect\relax
{\catcode`\|=\active
  \xdef\Braket{\protect\expandafter\noexpand\csname Braket \endcsname}
  \expandafter\gdef\csname Braket \endcsname#1{\begingroup
     \ifx\SavedDoubleVert\relax
       \let\SavedDoubleVert\|\let\|\BraDoubleVert
     \fi
     \mathcode`\|32768\let|\BraVert
     \left\langle{#1}\right\rangle\endgroup}
}
\def\BraVert{\@ifnextchar|{\|\@gobble}
     {\egroup\,\mid@vertical\,\bgroup}}
\def\BraDoubleVert{\egroup\,\mid@dblvertical\,\bgroup}
\let\SavedDoubleVert\relax

{\catcode`\|=\active
  \xdef\set{\protect\expandafter\noexpand\csname set \endcsname}
  \expandafter\gdef\csname set \endcsname#1{\mathinner
        {\lbrace\,{\mathcode`\|32768\let|\midvert #1}\,\rbrace}}
  \xdef\Set{\protect\expandafter\noexpand\csname Set \endcsname}
  \expandafter\gdef\csname Set \endcsname#1{\left\{%
     \ifx\SavedDoubleVert\relax \let\SavedDoubleVert\|\fi
     \:{\let\|\SetDoubleVert
     \mathcode`\|32768\let|\SetVert
     #1}\:\right\}}
}
\def\midvert{\egroup\mid\bgroup}
\def\SetVert{\@ifnextchar|{\|\@gobble}
    {\egroup\;\mid@vertical\;\bgroup}}
\def\SetDoubleVert{\egroup\;\mid@dblvertical\;\bgroup}

\begingroup
 \edef\@tempa{\meaning\middle}
 \edef\@tempb{\string\middle}
\expandafter \endgroup \ifx\@tempa\@tempb
 \def\mid@vertical{\middle|}
 \def\mid@dblvertical{\middle\SavedDoubleVert}
\else
 \def\mid@vertical{\mskip1mu\vrule\mskip1mu}
 \def\mid@dblvertical{\mskip1mu\vrule\mskip2.5mu\vrule\mskip1mu}
\fi

\usepackage{graphicx}
\usepackage[T1]{fontenc}
\usepackage[]{amsfonts}
\usepackage[]{amsmath}
\usepackage[]{rotating}
\usepackage[]{color}
\usepackage[]{float}
\usepackage[]{fancyhdr}
\usepackage[]{booktabs}
\usepackage[]{epsfig}
\usepackage{appendix}
\usepackage{amssymb}
\usepackage{psfrag}
\usepackage{setspace}

\begin{document}

\title{Quantum effect of inductance on geometric Cooper-pair transport}

\author{Juha Salmilehto}
\affiliation{COMP Centre of Excellence, Department of Applied Physics, Aalto University, P.O. Box 13500, FI-00076 AALTO, Finland}
\author{Mikko M\"ott\"onen}
\affiliation{COMP Centre of Excellence, Department of Applied Physics, Aalto University, P.O. Box 13500, FI-00076 AALTO, Finland}
\affiliation{Low Temperature Laboratory, Aalto University, P.O. Box 13500, FI-00076 AALTO, Finland}


\begin{abstract}

We introduce a model for a flux-assisted Cooper-pair pump, the sluice, which is used to study geometric charge transport. Our model allows for a nonvanishing loop inductance going beyond the usual treatment with an exact phase bias. We derive the device Hamiltonian and current operators for different elements of the system and calculate the pumped charge carried by the ground state in the adiabatic limit. We show that extending the model beyond the exact phase bias, has a weak but potentially non-negligible effect on the charge transport. This effect is observed to depend on the external flux bias. The adiabatic energy level and eigenstate structures are studied and the unusual features are explained. Finally, we discuss the requirements of adiabaticity.

\end{abstract}

\maketitle

\section{Introduction}

The methodology of constructing superconducting Josephson junction circuits\cite{rmp73/357} has provided a wondrous test-bed for many quantum phenomena. Perhaps one of the most intriguing applications is the possibility to study physics related to quantum phases\cite{GPIP}, most notably the Berry phase\cite{rmp82/1959, science318/1889}. Understanding the properties of these phases\cite{prsla392/45, prl51/2167, prl52/2111, prl58/1593, prl60/2339} has given rise to geometric quantum computing\cite{nature407/355, pla264/94} where the robustness against a certain type of noise is an inherent property of the computing scheme.

One of the most extensively studied Josephson devices related to geometric phases is the Cooper-pair sluice\cite{prl91/177003, apl90/082102}. It builds on the idea of implementing an array of Josephson junctions where single Cooper pairs are adiabatically transported in a controlled manner\cite{zpb85/349, prb60/R9931, prb64/172509, prb68/054510} establishing a link between the pumped charge and the Berry phase\cite{prb68/020502(R), prl95/256801}. The introduction of superconducting quantum interference devices (SQUIDs) operating as tunable junctions adds to the pump control\cite{prl91/177003}. Cooper-pair pumps based on such array structures have been theoretically shown to produce very accurate charge quantization\cite{prl98/127001, prl100/117001, prb77/144522} as well as having the potential for implementing non-Abelian structures\cite{prl100/027002, prb81/174506}. For the sluice, the connection between the pumped charge and the Berry phase has been shown both theoretically\cite{prb68/020502(R), prb73/214523} and experimentally\cite{prl100/177201}. Furthermore, the development of the sluice has given rise to proposals to implement geometric quantum computing featuring similar architectures\cite{pra82/052304, prb83/214518}. The sluice itself has been proposed to be used as a Landau--Zener--St\"uckelberg interferometer\cite{prl107/207002}.

During the past few years, theoretical effort has been put into modeling the sluice coupled to a dissipative environment. Using the usual reduced-density-operator methods\cite{API, tToOQS}, such a coupling leads to charge nonconservation. Recent papers have dealt with this problem using either superadiabatic bases\cite{prl105/030401, prb82/134517, pra82/062112, prb84/174507} or the Floquet theory\cite{prb83/214508, prb84/235140}. Both methods also have the potential for describing pumping when the driving is nonadiabatic, a regime recently studied experimentally~\cite{prb86/060502}. Furthermore, a current induced directly by the environment has been shown to emerge for certain types of noise operators\cite{prb85/024527} leading to a development of a conservation law for all operator currents in open quantum systems\cite{pra85/032110}. To date, all considerations of the sluice have assumed an exact phase bias. Namely, the inductance of a superconducting loop at the heart of the device has been assumed so small that the total gauge-invariant phase difference across all the weak links along the loop is fixed exactly by the penetrating magnetix flux. This assumption has simplified the theoretical analysis which has still been shown to provide predictions in agreement with the implemented experiments\cite{prl100/177201}. However, these experiments have not been carried out with very high precision and given that the sluice has been widely discussed as a candidate for a metrological current source, it is of utmost importance to study whether the finite realistic loop inductance plays a role here.

In this paper, we introduce a theoretical model for the sluice which allows the inclusion of a nonvanishing loop inductance. The derived Hamiltonian accounts for the emerging additional quantum degrees of freedom. The current operators for the superconducting island and the effective inductor are derived from first principles. We calculate the geometric charge carried by the adiabatic ground state emphasizing the features due to the finite loop inductance and show that the results given by the model with vanishing inductance are valid to a rather good degree with experimentally relevant parameters. Variation of the additional parameters included in our model is shown to weakly affect the pumped charge. We study the instantaneous eigenenergies and eigenstates to point out the differences between our more general model and the case of the exact phase bias.  Finally, we consider the adiabaticity criteria for the system.

The paper is organized as follows. In Sec.~\ref{subsec:H}, we introduce our model for the sluice and derive the Hamiltonian for it. In Sec.~\ref{subsec:currents}, we derive current operators for the different parts of  the system without resorting to operator derivatives. In Sec.~\ref{subsec:perturbative}, we study analytically the transferred geometric and dynamic charges in first order perturbation theory. The charges are found equal to the ones obtained using the original model. In Sec.~\ref{subsec:numerical}, we perform a numerical analysis of both the system and its charge transport features in the adiabatic limit. The geometric charge carried by the ground state is shown to differ slightly from the original model. The features of the energy level structure and eigenstates are explained and the requirements of adiabaticity are discussed briefly. We conclude the paper in Sec.~\ref{sec:conclusions}.

\section{Model device}

\subsection{Hamiltonian} \label{subsec:H}

\begin{figure}
\includegraphics[width=8.5cm]{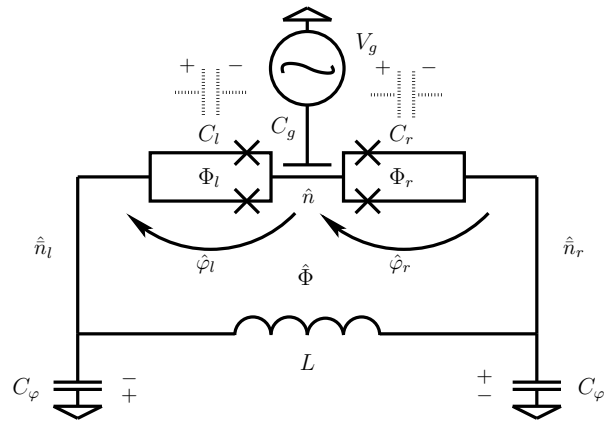}
\caption{Circuit diagram of the Cooper pair sluice. The external fluxes threading the left and right SQUIDs are denoted by $\Phi_l$ and $\Phi_r$, respectively, and $\hat{\Phi}$ denotes the operator for the total flux threading the large superconducting loop. The operators for the phase differences over the SQUIDs are marked by $\hat{\varphi}_l$ and $\hat{\varphi}_r$ and they are defined in the direction specified by the arrows in the figure. The internal capacitances of the SQUIDs are $C_l$ and $C_r$, the loop is connected to the ground terminal via two identical capacitors with capacitances $C_{\varphi}$, and $L = L_G + L_K$ denotes the total inductance of the loop comprising of both the geometric $L_G$ and the kinetic $L_K$ contributions. The gate capacitance $C_g$ is used to manipulate the island charge with the gate voltage $V_g$. The operators $\hat{n}$, $\hat{\bar{n}}_l$, and $\hat{\bar{n}}_r$ are for the Cooper-pair numbers of the corresponding islands. Note that the superconducting islands described by $\hat{\bar{n}}_l$ and $\hat{\bar{n}}_r$ are inductively coupled. The choice of the charge orientation at the capacitors for positive voltages across the capacitor plates is marked by $\{ +,- \}$ including the effective junction capacitors denoted by dashed lines.}
\label{fig:sluice}
\end{figure}

The device structure we use is an adaptation of the original model used for the Cooper pair sluice\cite{prl91/177003}. The original model consists of a superconducting island isolated by two SQUIDs positioned in a large superconducting loop [see Fig.~\ref{fig:sluice} for our model resembling the original structure with respect to the device operation]. The island is capacitively coupled to a gate terminal used to manipulate its charge state. The leads are taken to have capacitive couplings to a ground terminal. The self-inductances of the SQUID loops are assumed negligible so that they act as tunable Josephson junctions whose Josephson energies can be controlled with the external magnetic fluxes threading the loops. Phase-biasing is typically introduced for the device by assuming that the total inductance of the large loop is negligible implying that the total superconducting phase difference over all the weak links in the loop is a real number exactly determined by the penetrating magnetic flux. This assumption allows one to determine the system dynamics using only the island charge basis.

We improve on the existing model by introducing an inductive element allowing us to depict, among other inductance sources, the self-inductance of the loop\footnote{We assume that the effect of the self-inductance of the loop is dominant in the transported charges allowing us to neglect the self-inductance of the SQUIDs. This is because the self-inductance of the loop directly affects the phase bias whereas the self-inductance of the SQUIDs induces nonideality in the SQUID operation mostly accounted for by the residual values of the Josephson energy scales. On the other hand, we aim to study cases where the loop inductance obtains significantly large values greatly exceeding the values corresponding to the experimental realization of the SQUIDs.}. Ground couplings are established from both sides of the element using two identical shunt capacitors. The resulting device is shown in Fig.~\ref{fig:sluice}. 
The Hamiltonian for the device can be written as [see Appendix~\ref{sec:const_ham} for details]
\begin{equation}
\begin{split}
\hat{H} &= E_C(\hat{n}-n_g)^2 + E_{\varphi} \hat{\bar{n}}^2 + \hat{H}_{\mathrm{FIN}} + \frac{E_L}{2}(\hat{\varphi}-\varphi_0)^2 \\ &- E_{Jr}\cos(\hat{\phi}+\hat{\varphi}/2) - E_{Jl}\cos(\hat{\phi}-\hat{\varphi}/2),
\end{split}
\label{eq:H_orig}
\end{equation}
where $\hat{\phi}$ is the island phase operator, $\hat{\varphi}$ is the operator for the total phase difference over the SQUIDs, $\hat{n}$ is the Coope- pair number (CPN) operator for the island and $\hat{\bar{n}} = (\hat{\bar{n}}_l - \hat{\bar{n}}_r)/2$ is the Cooper-pair number operator for the feed; that is, it describes the average charge imbalance between the left and right leads. The CPN operators for the left and right leads are given by $\hat{\bar{n}}_l$ and $\hat{\bar{n}}_r$, respectively. To allow for a form resembling the case of the exact phase bias, we omitted explicitly writing all identity operators but will reestablish their use in the following presentations of the Hamiltonian. Here, $E_C$ is the charging energy scale for the island [see Eq.~(\ref{eq:charging_normal})], $E_{\varphi}$ is the charging energy scale for the feed [see Eq.~(\ref{eq:charging_normal})], $E_L =  \frac{1}{L} \left( \frac{\Phi_0}{2\pi} \right)^2$ is the inductive energy scale due to the finite total inductance $L$ of the loop, $\hat{H}_{\mathrm{FIN}}$ is an additional charging Hamiltonian stemming from the derivation [see Eqs.~(\ref{eq:HFIN}) and (\ref{eq:charging_HFIN})], $\varphi_0 = \frac{2\pi \Phi_{\mathrm{ext}}}{\Phi_0}$ is related the external magnetic flux $\Phi_{\mathrm{ext}}$, $n_g = \frac{V_gC_g}{2e}$ is the normalized gate charge, and $E_{Jl},E_{Jl}$ are the Josephson energies of the tunable junctions. The flux quantum is denoted by $\Phi_0 \approx 2.07$ fWb, the elementary charge is denoted by $e$ and $L = L_G + L_K$, where $L_G$ is the total geometric inductance of the loop and $L_K$ is the corresponding kinetic inductance. The derivation of the Hamiltonian and the exact definitions of all the quantities are given in Appendix~\ref{sec:const_ham}. 

We have written the Hamiltonian using two pairs of conjugate variables; that is, $(\hat{n},\hat{\phi})$ and $(\hat{\bar{n}},\hat{\varphi})$. Especially, the definition of the feed CPN operator allows us to treat the inductive part of the Hamiltonian. We identify 
\begin{equation}
\begin{split}
\hat{H}_{\varphi} &= E_{\varphi} \hat{\bar{n}}^2 + \frac{E_L}{2}(\hat{\varphi}-\varphi_0)^2 \\ &= -E_{\varphi} \frac{\partial^2}{\partial \hat{\varphi}^2} + \frac{E_L}{2}(\hat{\varphi}-\varphi_0)^2,
\end{split}
\end{equation}
as being the Hamiltonian for a phase-shifted quantum harmonic oscillator, for which the effective mass is $m=\hbar^2/(2E_{\varphi})$ and the effective angular frequency is $\omega = \sqrt{2E_{\varphi}E_L}/\hbar$. If we denote $d_0 = (2E_{\varphi}/E_L)^{1/4}$, the corresponding eigenproblem $\hat{H}_{\varphi}\ket{\Psi_k} = E_k \ket{\Psi_k}$ defines the Fock states for the feed
\begin{equation}
\begin{split}
\braket{\varphi|\Psi_k} &= (2^k k! \sqrt{\pi} d_0)^{-1/2} \exp \left( -\frac{1}{2} \left( \frac{\varphi-\varphi_0}{d_0} \right)^2 \right) \\ &\times H_k \left( \frac{\varphi - \varphi_0}{d_0} \right),
\end{split}
\label{eq:Vphi}
\end{equation}
where $k$ is any non-negative integer and $H_k$ is the $k$th Hermite polynomial. The corresponding energies are given by
\begin{equation}
\begin{split}
E_k &= \sqrt{2E_{\varphi}E_L} \left( k+\frac{1}{2} \right).
\end{split}
\label{eq:Ephi}
\end{equation}
Notice that the canonical commutation relation is $[\hat{\varphi},\hat{\bar{n}}] = i$ implying that the effective momentum operator is $\hbar \hat{\bar{n}}$. Defining the relevant bosonic creation and annihilation operators for the feed Fock states as $\hat{a}^{\dagger}$ and $\hat{a}$, respectively, allows us to write the total phase difference operator as
\begin{equation}
\begin{split}
\hat{\varphi}  = \left( \frac{E_{\varphi}}{2E_L} \right)^{1/4} (\hat{a}+\hat{a}^{\dagger}) + \varphi_0.
\end{split}
\label{eq:varphi}
\end{equation}

Using the above-mentioned notation, the Hamiltonian can be written as
\begin{equation}
\begin{split}
\hat{H} &= \hat{H}_n \otimes \hat{\mathbb{I}}_{\varphi} + \hat{\mathbb{I}}_n \otimes \hat{H}_{\varphi} + \hat{H}_{\mathrm{FIN}} \\ &- \frac{E_{Jr}}{2} \{ \hat{n}_+ \otimes e^{i\varphi_0 /2} \hat{D}(ig_0) + \hat{n}_- \otimes e^{-i\varphi_0 /2} \hat{D}(-ig_0) \} \\ &- \frac{E_{Jl}}{2} \{ \hat{n}_+ \otimes e^{-i\varphi_0 /2} \hat{D}(-ig_0) + \hat{n}_- \otimes e^{i\varphi_0 /2} \hat{D}(ig_0) \},
\end{split}
\label{eq:H_mod}
\end{equation}
where $\hat{H}_n = E_C(\hat{n}-n_g)^2$, $\hat{\mathbb{I}}_n$ is the identity operator in the island CPN basis, $\hat{\mathbb{I}}_{\varphi}$ is the identity operator in the feed Fock basis, $\hat{n}_+ = \exp [i\hat{\phi}]$ ($\hat{n}_- = \exp [-i\hat{\phi}]$) is the raising (lowering) operator in the island charge basis and $\hat{D}(ig_0) = \exp[ig_0(\hat{a}+\hat{a}^{\dagger})]$, where 
\begin{equation}
g_0=\left( \frac{E_{\varphi}}{32E_L} \right)^{1/4}.
\end{equation}
The operator $\hat{D}(ig_0)$ is the displacement operator for one mode in the quantum phase space with $ig_0$ being the displacement. This notation is typically utilized in quantum optics. Finally, we rewrite the Hamiltonian as
\begin{equation}
\begin{split}
\hat{H} &= \hat{H}_S \otimes \hat{\mathbb{I}}_{\varphi} + \hat{\mathbb{I}}_n \otimes \hat{H}_{\varphi} + \hat{H}_{\mathrm{FIN}} \\ &+ \hat{H}_B \otimes [\hat{D}(ig_0)-\hat{\mathbb{I}}_{\varphi}] + \hat{H}_B^{\dagger} \otimes [\hat{D}^{\dagger}(ig_0)-\hat{\mathbb{I}}_{\varphi}],
\end{split}
\label{eq:H_final}
\end{equation}
where $\hat{H}_S = E_C(\hat{n}-n_g)^2 - E_{Jr} \cos (\hat{\phi}+\varphi_0/2) - E_{Jl} \cos (\hat{\phi}-\varphi_0/2)$ is the Hamiltonian for the original model with exact phase bias and $\hat{H}_B = -\frac{e^{i\varphi_0/2}}{2} [E_{Jr}\hat{n}_+ + E_{Jl}\hat{n}_-]$. Note that if the total inductance is negligible ($L\to 0$, $E_L\to\infty$), the system dynamics are accurately described by the original Hamiltonian acting only on the island degrees of freedom.

\subsection{System currents} \label{subsec:currents}

To study the currents flowing in the device, we describe the temporal change of the expectation value of the feed charge operator using the corresponding current operator $\hat{I} = -\frac{2ei}{\hbar} [\hat{H},\hat{\bar{n}}]$, where we exploited the usual current identity applicable for any arbitrary charge operator that does not depend on time explicitly\cite{prb73/214523, prb84/174507}. We thus proceed to write the feed CPN operator as
\begin{equation}
\begin{split}
\hat{\bar{n}} = -i\frac{1}{4g_0}(\hat{a}-\hat{a}^{\dagger}),
\end{split}
\label{eq:barn}
\end{equation}
which allows us to write the general form for the current operator as
\begin{equation}
\begin{split}
\hat{I} = -\frac{e}{2\hbar g_0} [\hat{H}, \hat{\mathbb{I}}_n \otimes (\hat{a}-\hat{a}^{\dagger})].
\end{split}
\label{eq:I_orig}
\end{equation}
Exploiting the Hamiltonian derived in the previous section and utilizing some of the properties of the displacement operator familiar from the coherent state formalism\cite{QO}, the current operator becomes
\begin{equation}
\begin{split}
\hat{I} &= \frac{ei}{\hbar} [\hat{H}_B \otimes \hat{D}(ig_0)-\hat{H}_B^{\dagger} \otimes \hat{D}^{\dagger}(ig_0)] \\ &+ \frac{2e}{\hbar} \left( \frac{E_{\varphi}E_L^3}{2} \right)^{1/4} \hat{\mathbb{I}}_n \otimes (\hat{a}+\hat{a}^{\dagger}).
\end{split}
\label{eq:I_final}
\end{equation}
Note that the first and third terms of the Hamiltonian in Eq.~(\ref{eq:H_final}) do not contribute anything to the current operator due to commutation. If the total inductance is negligible, the first term on the right-hand side in Eq.~(\ref{eq:I_final}) yields the current operator used in the original model.

The current operator in Eq.~(\ref{eq:I_final}) can be interpreted in terms of currents flowing in the system. The operator $\hat{I}$ describes the temporal change in the charge imbalance and, hence, the transfer of charge between the shunt capacitors or the effective junction capacitors. Due to the device structure, the charge can be transferred via two separate routes: through the inductor or through the island, and the total transfer must be described by $\hat{I}$.

\begin{figure}
\includegraphics[width=8.5cm]{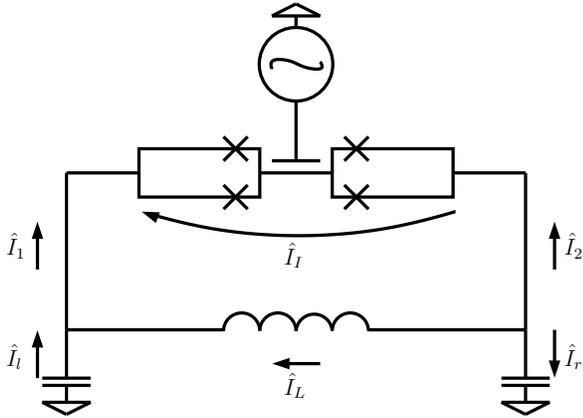}
\caption{Schematic diagram of the electric currents for the model used to analyze the Cooper-pair sluice. See text for further explanation of the current operators.}
\label{fig:current_schematic}
\end{figure}

To quantify the argument given above, let us present the current schematic for the sluice in Fig.~\ref{fig:current_schematic}. 
In the figure, $\hat{I}_I$ is the operator for the average current tunneling through the junctions, $\hat{I}_L$ is the operator for the current flowing through the inductor and $\hat{I}_1$ ($\hat{I}_2$) describes the current flowing to the junction system from the left (right). Furthermore, we denote the operator for the current leaving the left shunt capacitor by $\hat{I}_l$ and for the current arriving to the right shunt capacitor by $\hat{I}_r$. By the definition of the feed charge operator presented in Appendix~\ref{sec:const_ham}, we have $\hat{I} = \hat{I}_1+\hat{I}_I-\hat{I}_l$, where we use the fact that the system of islands is capacitively isolated from any external current sources. By current conservation, $\hat{I}_1 = \hat{I}_l+\hat{I}_L$ and, hence, $\hat{I} = \hat{I}_L+\hat{I}_I$. Thus, the simple argument given above about the separation of $\hat{I}$ into different current operators for the system has been shown valid. In addition, we have $-\hat{I}_r = \hat{I}_L+\hat{I}_2$. If the current arriving to the island equals the average current tunneling through the junctions $\hat{I}_2 = \hat{I}_I$, then the conservation of total charge $\hat{I}_r = \hat{I}_l$ implies that $\hat{I} = - \hat{I}_r = -\hat{I}_l$ indicating that the current operator describes discharging (charging) of the right (left) shunt capacitor.

Identifying the current operators for the inductive element and for the tunneling current across the island is straightforward. As described in Appendix A, the inductive current can be derived by studying the operator for the total flux penetrating the loop $\hat{\Phi} = \Phi_{\textrm{ext}} + L_G\hat{I}_L$ and its relation to the operator for the total phase difference over the SQUIDs $\frac{2\pi \hat{\Phi}}{\Phi_0} = \hat{\varphi} - \frac{2e}{\hbar} L_K\hat{I}_L$. Rewriting the inductive current operator given in Eq.~(\ref{eq:IL_basic}) yields
\begin{equation}
\begin{split}
\hat{I}_L = \frac{2e}{\hbar} E_L (\hat{\varphi}-\varphi_0) = \frac{2e}{\hbar} \frac{\partial \hat{H}_{\mathrm{ind}}}{\partial \hat{\varphi}},
\end{split}
\label{eq:I_L_1}
\end{equation}
where the last form is just to show that the usual manner of defining via operator derivates is also applicable. Formulating this current operator in terms of the creation and annihilation operators allows us to write
\begin{equation}
\begin{split}
\hat{I}_L &= \frac{2eE_L}{\hbar} [2g_0(\hat{a}+\hat{a}^{\dagger})+\varphi_0-\varphi_0] \\ &= \frac{2e}{\hbar} \left( \frac{E_{\varphi}E_L^3}{2} \right)^{1/4} \hat{\mathbb{I}}_n \otimes (\hat{a}+\hat{a}^{\dagger}),
\end{split}
\label{eq:I_L_2}
\end{equation}
which can be identified as the second term in $\hat{I}$ in Eq.~(\ref{eq:I_final}). Using the operator derivative, we could also identify the first term in Eq.~(\ref{eq:I_final}) as stemming from the Josephson part of the Hamiltonian and, hence, describing the charge passing through the island. However, due to the above-derived separation of currents, such application of operator derivatives is not necessary in order to define the island current operator.

In the adiabatic limit, there is no voltage over the inductor so that the potentials on the left and right islands are equal. Thus, there is no potential difference that would induce a change in the charge imbalance between the left and right leads, and hence $\braket{\hat{I}}=0$ at all times; that is, there is no net current flowing in or out of the shunt capacitors. In this case, we can define the charge circulating in the loop to be described by either $\hat{I}_L$ or $\hat{I}_I$.

\section{Analysis} \label{sec:results}

Using the assumption of negligible loop inductance, the geometric charge transferred adiabatically through the sluice during one parameter cycle has been theoretically studied in several references~\cite{prl91/177003, prb73/214523, prl105/030401, prb82/134517, prb83/214508, prb84/174507, prb84/235140, prb85/024527}. However, a model enabling one to use a non-negligible inductance for such transfer has not been studied previously. We write the geometric charge transferred by the $m$th adiabatic state in the usual manner\cite{prb60/R9931, prl91/177003, prb73/214523}
\begin{equation}
\begin{split}
Q_{(I/L)}^{(G,m)} = &2\hbar \Im m \int_{\gamma} \sum_{k\neq m}  \frac{\braket{m(\vec{q})|\hat{I}_{(I/L)}(\vec{q})|k(\vec{q})}}{E_{m}(\vec{q})-E_{k}(\vec{q})} \\ &\times \braket{k(\vec{q})|\nabla_{\vec{q}}|m(\vec{q})} \cdot d\vec{q},
\end{split}
\label{eq:Q_G}
\end{equation}
where $\ket{i(\vec{q})}$ is the $i$th instantaneous eigenstate of the Hamiltonian in Eq.~(\ref{eq:H_final}) and $E_i(\vec{q})$ is the corresponding eigenenergy. We present everything as a function of the point in the control parameter space denoted by the vector $\vec{q}$ whose scalar components are the external control parameters. The integral is over a contour $\gamma$ in the parameter space. In the case of cyclic evolution, the contour is closed and the pumped charge is related to the Berry phase obtained during the time evolution\cite{prb68/020502(R), prb73/214523, prl100/177201}. The pumped charge is fully determined by the contour and, hence, it does not depend on the pumping speed. For analytical calculations, we also define the dynamic charge carried by the supercurrent as
\begin{equation}
\begin{split}
Q_{(I/L)}^{(D,m)} &= \int_0^{1/f} dt \braket{m(t)|\hat{I}_{(I/L)}(t)|m(t)},
\end{split}
\label{eq:Q_D}
\end{equation}
where the pumping frequency is denoted by $f$.

In the following, we assume that the island is capacitively symmetric $C_l = C_r = C_J$ and that $C_g \ll C_{\varphi}$. According to Appendix~\ref{sec:const_ham}, $\hat{H}_{\mathrm{FIN}}$ can consequently be neglected in the total Hamiltonian. Furthermore, we assume that the external control parameters are manipulated slowly enough for the evolution to remain adiabatic.

\subsection{Perturbative approach in the lowest order in $g_0$} \label{subsec:perturbative}

Perturbation theory can be applied in the lowest order in $g_0$ to obtain analytical information on the properties of the system. Decomposing the displacement operator yields
\begin{equation}
\begin{split}
\hat{D}(ig_0) = \hat{\mathbb{I}}_{\varphi} + ig_0(\hat{a}+\hat{a}^{\dagger}) + O(g_0^2).
\end{split}
\label{eq:D_decomp}
\end{equation}
If $E_{\varphi} \ll E_L$, $g_0$ is small and the linear term suffices so that the Hamiltonian in Eq.~(\ref{eq:H_final}) is accurately approximated by
\begin{equation}
\begin{split}
\hat{H}  \approx \hat{H}_S \otimes \hat{\mathbb{I}}_{\varphi} + \hat{\mathbb{I}}_n \otimes \hat{H}_{\varphi} + i g_0 (\hat{H}_B-\hat{H}_B^{\dagger}) \otimes (\hat{a}+\hat{a}^{\dagger}).
\end{split}
\label{eq:H_firstorder}
\end{equation}
Similarly, taking only the linear term in Eq.~(\ref{eq:I_final}) yields
\begin{equation}
\begin{split}
\hat{I} &= \hat{I}_S \otimes \hat{\mathbb{I}}_{\varphi} - g_0 \frac{e}{\hbar} (\hat{H}_B + \hat{H}_B^{\dagger}) \otimes (\hat{a} + \hat{a}^{\dagger}) \\ &+ \frac{2e}{\hbar} \left( \frac{E_{\varphi}E_L^3}{2} \right)^{1/4} \hat{\mathbb{I}}_n \otimes (\hat{a}+\hat{a}^{\dagger}),
\end{split}
\label{eq:I_firstorder}
\end{equation}
where we have identified $\hat{I}_S = \frac{ei}{\hbar} (\hat{H}_B-\hat{H}_B^{\dagger})$ as the current operator in the original model. For future reference, we denote the first term on the right-hand side as the zeroth order island current operator $\hat{I}_I^{(0)}$ and the second term as the first order island current operator $\hat{I}_I^{(1)}$.

After an analytical calculation, it can be shown that no first-order perturbative correction in $g_0$ emerges to the dynamic or geometric currents across the island compared with the original model. Furthermore, the corresponding currents through the inductor can be shown to be equal but of opposite sign up to this order. This observation agrees with our previous assumption that the total temporal change of the expectation value of the feed CPN operator is zero in the adiabatic limit. See Appendix~\ref{sec:firstorder} for details of the perturbative calculation. Exploiting higher-order perturbation theory would require a higher-order expansion of the displacement operator in Eq.~(\ref{eq:D_decomp}) leading to an increasingly more difficult analytical analysis. Furthermore, we wish to study a region where $g_0$ is significantly large so that the perturbative approach is no longer practical. Hence, we resort to numerical analysis below to study the current contribution arising from the finite loop inductance.

\subsection{Numerical analysis} \label{subsec:numerical}

We begin our analysis by discussing the physical energy scales. Assuming that $C_g$ is nearly negligible, the charging energy scales are determined by $E_C \approx e^2/C_J$ and $E_{\varphi} \approx 4e^2/(C_{\varphi}+C_J)$ [see Appendix~\ref{sec:const_ham}]. To establish a comparison with recent studies~\cite{prl105/030401, prb82/134517, prb83/214508, prb84/174507, prb84/235140, prb85/024527}, we assume that $E_C/k_B = 1$ K, where $k_B$ is the Boltzmann constant. This defines the junction capacitances $C_J = C_l = C_r$ to be of the order of femtofarads. We study a range from 0.05 to 500 fF in the shunt capacitances $C_{\varphi}$. We can assume, for example, that the shunt capacitances are primarily due to capacitive couplings to a bottom ground terminal via bonding pads. If the device is constructed on a 500 $\mu$m thick silicon wafer with square-shaped pads, the equivalent range in the length of the side of the pads is from 15 $\mu$m to 1.5 mm yielding experimentally relevant values. Using the above-mentioned capacitances, the maximum feed energy scale is limited by $C_J$ so that $E_{\varphi}$ decreases from $4E_C$ to $10^{-2}E_C$ with increasing pad capacitance. 

The inductive energy scale is determined by $E_L = \frac{1}{L} \left( \frac{\Phi_0}{2\pi} \right)^2$ [see Appendix~\ref{sec:const_ham}]. We approximate the geometric inductance of the wire by $L_G \approx \mu l$, where $\mu$ is the permeability of the wire material and $l$ is the length of the large superconducting loop. The kinetic inductance can be written in the BCS formalism as $L_K = \left( \frac{l}{w} \right) \frac{R_{\mathrm{sq}}h}{2\pi^2\Delta} \frac{1}{\tanh \left( \Delta / 2k_B T \right)}$, where $w$ is the width of the superconducting wire, $R_{\mathrm{sq}}$ is the sheet resistance of the wire in  the non-superconducting state, $\Delta$ is the superconductor energy gap and $T$ is the wire temperature\cite{nanotech21/445202, ItS}. We assume that $T \ll T_c$, where $T_c$ is the critical temperature of the superconductor, and that the wire is made of aluminum such that $w$ is of the order of a micrometer\cite{prl100/177201}. In addition,  we assume that the thickness of the wire is of the order of tens of nanometers or larger limiting the kinetic inductance to be of the same order than the geometric one or smaller. The feasible length scales for the loop range between 10 $\mu$m and 1 cm corresponding to values between $10^3E_C$ and $E_C$ in the inductive energy scale. 

The standard control parameter cycle utilized in Refs.~\onlinecite{prl105/030401},~\onlinecite{prb82/134517},~\onlinecite{prb83/214508, prb84/174507, prb84/235140}, and~\onlinecite{prb85/024527} is adopted and the device is operated in the charging regime $E_C \gg \max \{E_{Jl},E_{Jr} \}$. The parameter cycle is presented in Fig.~\ref{fig:cycle} [see Appendix~\ref{sec:cycle} for a detailed description].
\begin{figure}
\includegraphics[width=8.5cm]{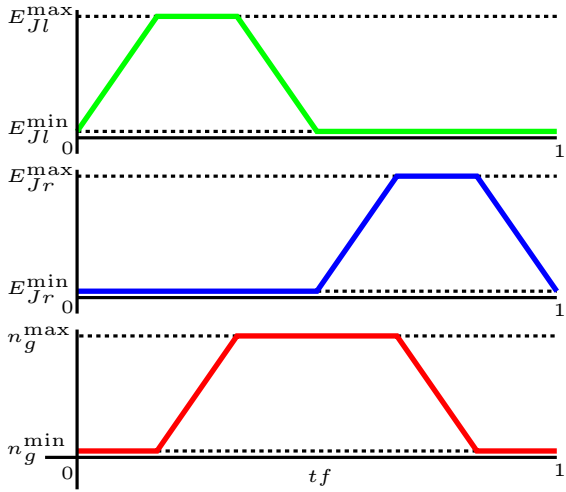}
\caption{(Color online) Time dependence of the control parameters $E_{Jl}$, $E_{Jr}$, and $n_g$ during a Cooper-pair pumping cycle.}
\label{fig:cycle}
\end{figure}
Furthermore, the normalized gate charge $n_g$ is manipulated near 1/2 so that the required island states are limited near this value. In the original model, using only the two closest island CPN states $\ket{n=0}$ and $\ket{n=1}$ was sufficient. We conduct our simulations using a basis with an even number $N_i$ of island CPN states such that the basis consists of $N_i/2$ pairs of states $\ket{n=1-i}$ and $\ket{n=i}$, where $i=1,\dots,N_i/2$, and $N_f$ feed Fock states with the lowest energies.

\begin{figure*}
\includegraphics[width=18cm]{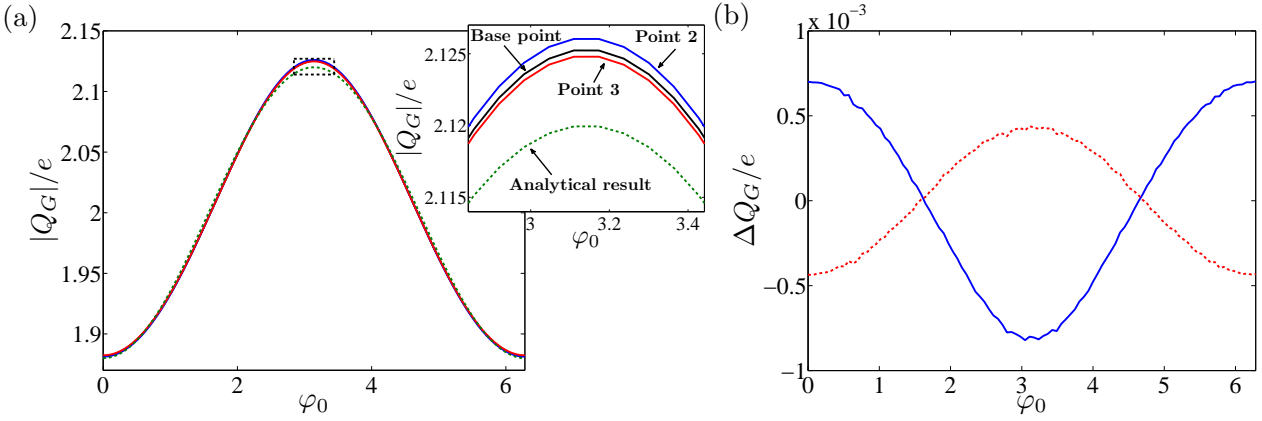}
\caption{(Color online) (a) Pumped charges and (b) charge differences as a function of the scaled external magnetic flux bias $\varphi_0 = \frac{2\pi \Phi_{\mathrm{ext}}}{\Phi_0}$. We use $E_C/k_B = 1$ K, $E_{Jr}^{\mathrm{max}}/E_C = E_{Jl}^{\mathrm{max}}/E_C = 0.1$, $E_{Jr}^{\mathrm{min}}/E_C = E_{Jl}^{\mathrm{min}}/E_C = 0.003$, $n_g^{\mathrm{max}} = 0.8$, $n_g^{\mathrm{min}} = 0.2$. The parameter points in (a) are $E_L = 10^2E_C$ and $E_{\varphi}=E_C$ (base point), $E_L = E_C$ and $E_{\varphi}=10^{-2}E_C$ (point 2), and $E_L = E_C$ and $E_{\varphi} = 4E_C$ (point 3). The dashed line depicts the approximate analytical result derived for the original model and the dashed rectangle indicates the area of the inset. In (b), we give the geometric charge differences between the base point and point 2, and between the base point and point 3 (from bottom to top at $\varphi_0 = \pi$). For the simulations, we use $N_i = 2$ and $N_f = 25$. }
\label{fig:run456g_QG}
\end{figure*}

According to Eq.~(\ref{eq:H_final}), the interesting features are characterized by two factors: the energy gap $\Delta_f \approx \sqrt{2E_{\varphi}E_L}$ between the states in different Fock subspaces and $g_0 = \left( \frac{E_{\varphi}}{32E_L} \right)^{1/4}$ dictating the strength of the coupling between the island and feed degrees of freedom. To assess the implications of these features with respect to the original model, we consider three different points in the $(E_L,E_{\varphi})$-space. Firstly, we establish a base point corresponding roughly to the energy scales used in the experiments of Ref.~\onlinecite{prl100/177201} using $E_L = 10^2E_C$ and $E_{\varphi}=E_C$. Secondly, we depict the point of smallest $\Delta_f$ within the feasible energy scales using $E_L = E_C$ and $E_{\varphi}=10^{-2}E_C$. Thirdly, we illustrate the point of largest $g_0$ using $E_L = E_C$ and $E_{\varphi} = 4E_C$. We perform the simulations for all values of the flux bias and retrieve a pumped charge close to the usual cosine dependence $Q_G = 2e ( 1 - 2 \epsilon \cos\varphi_0 ) + O(\epsilon^2)$, where $\epsilon = E_{Jk}^{\mathrm{min}} / E_{Jk}^{\mathrm{max}}$, $k \in \{l,r\}$~\cite{prb73/214523, prl100/177201}. Identical SQUIDs are assumed and we exploit the parameter cycle given in Appendix~\ref{sec:cycle}. The flux bias dependence of the pumped charge at the three selected parameter points is given in Fig.~\ref{fig:run456g_QG}(a) and the charge differences between the base point and the other two points in Fig.~\ref{fig:run456g_QG}(b).
The charge transferred at the base point is equal to the one given by the original model within the accuracy of the simulations. Note however that the accuracy of both simulations is orders of magnitude lower than the metrological accuracy $10^{-8}$. Furthermore, the charges transferred through the island and through the inductor were found equal in all the simulations validating the assumptions made about the adiabatic limit in Sec.~\ref{subsec:currents}.

Near the phase values $\varphi_0 = \pi/2 + N\pi$, where $N \in \mathbb{Z}$, the pumped charge is robust against variation of the additional parameters included in our model whereas at $\varphi_0 = N\pi$, the absolute difference between the charges reaches its local maxima; see Fig.~\ref{fig:run456g_QG}(b). Furthermore, averaging over the flux bias yields a vanishing charge difference. Both of these features are likely caused by the shape and symmetry of the parameter cycle. The two quantum degrees of freedom in our model are coupled by the tunneling term in the Hamiltonian. A crude estimate would thus state that the feed degree of freedom acts as if the original model obtains a reshaping of the parameter cycle for the SQUIDs explaining the vanishing effect on the pumped charge at approximately $\varphi_0 = \pi/2 + N\pi$, the same point where the vanishing effect of the residual Josephson energies appeared in the original analytical model. Note that treating the inductance classically within the original model fails to capture the above-mentioned dependence on the external flux\footnote{Assume that the total inductance induces a shift in the phase bias $\frac{2\pi}{\Phi_0}LI$ where $I$ is the total circulating current. The leakage supercurrent $I_S$ typically dominates over the geometric current and we can approximate $I \approx I_S$ which is nearly constant in time. Thus, the analytical formula for the original model yields a dominant correction to the transferred geometric charge proportional to $\sin \varphi_0 \times \frac{2\pi}{\Phi_0}LI_S$. This result indicates a difference in the external flux dependence of the geometric charge between the classical treatment of the inductance within the original model and our model. Thus, applying the quantum treatment is necessary to correctly model the effect of the inductance.}. The observed weak influence of altering the parameters is especially remarkable since, according to Eq.~(\ref{eq:Q_G}), the geometric charge is not affected only by changes in the state carrying it but also by changes in all the other states. 

\begin{figure*}
\includegraphics[width=18cm]{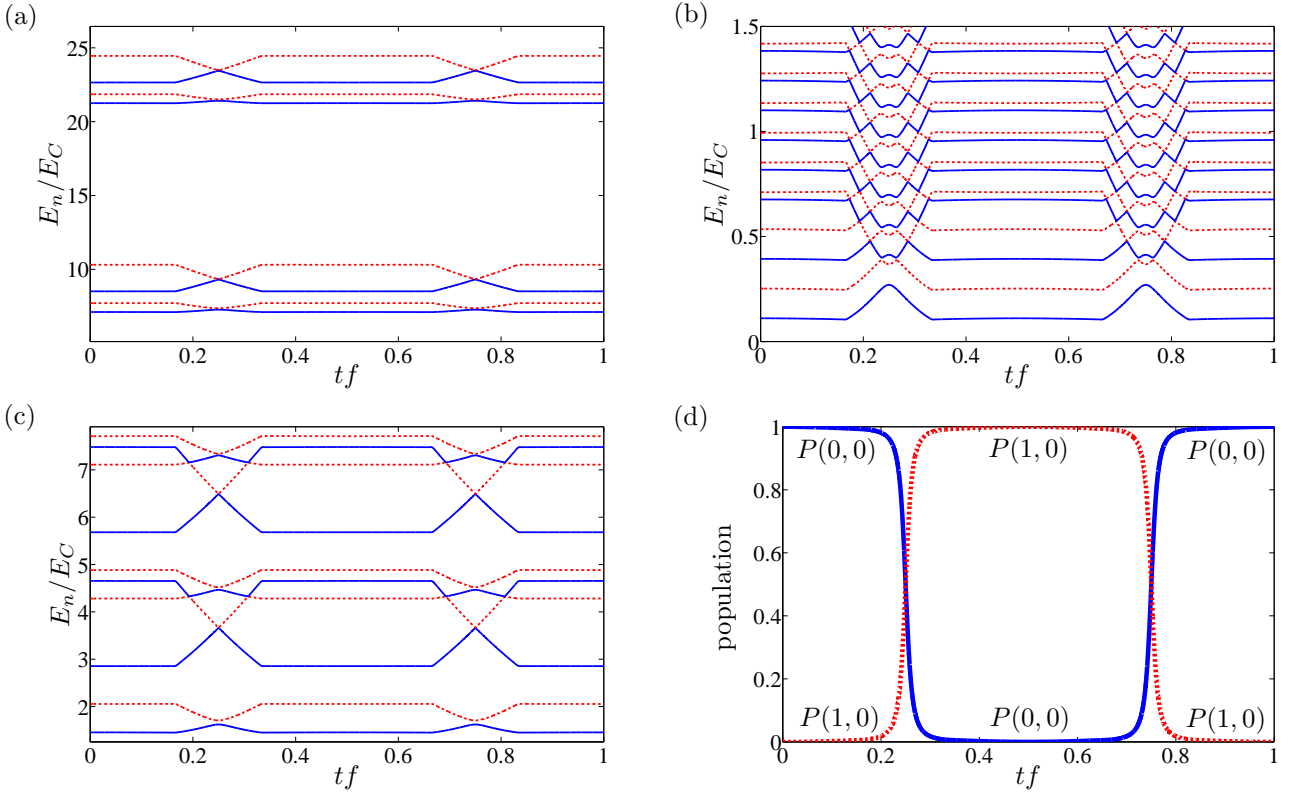}
\caption{(Color online) Energy level structures and ground-state probability distributions of the sluice as a function of scaled time $tf$, where $t$ is time and $f$ is the pumping frequency. Lowest instantaneous energy levels for (a) $E_L = 10^2E_C$ and $E_{\varphi}=E_C$, (b) $E_L = E_C$ and $E_{\varphi}=10^{-2}E_C$, and (c) $E_L = E_C$ and $E_{\varphi} = 4E_C$ during a pumping cycle detailed in Appendix~\ref{sec:cycle}. The penetrating magnetic flux is selected so that $\varphi_0 = \pi/2$ and the other parameters are equal to the ones used in Fig.~\ref{fig:run456g_QG}. The distribution of the ground-state populations $P(n,k)$ on different basis states [see text for further details] is given in panel (d). The differences in the distributions between the different parameter points are smaller than the line width in the figure and, hence, we only present the distribution for the base point for clarity. We use $N_i = 4$ and $N_f = 25$.}
\label{fig:EandV}
\end{figure*}

\begin{figure*}
\includegraphics[width=17cm]{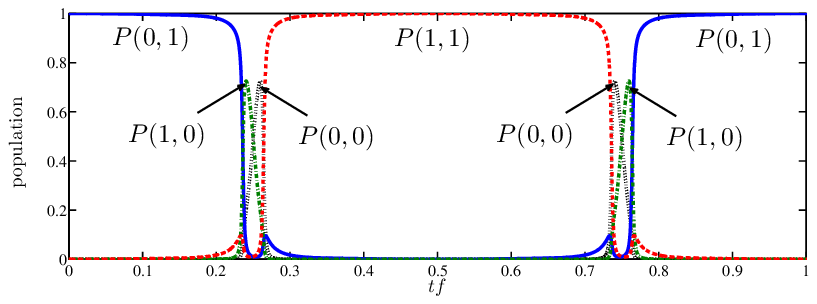}
\caption{(Color online) Distribution of the first-excited-state populations $P(n,k)$ on different basis states [see text for further details] for $E_L = E_C$ and $E_{\varphi}=10^{-2}E_C$ as a function of scaled time $tf$, where $t$ is time and $f$ is the pumping frequency. Other physical parameters are the same as in Fig.~\ref{fig:run456g_QG} with $\varphi_0 = \pi/2$. For the simulations, we use $N_i = 4$ and $N_f = 25$.}
\label{fig:excited5d}
\end{figure*}

We study the instantaneous energy level structure and eigenstates emerging from our model focusing near the point of maximum robustness; that is, we select the external magnetic field to yield $\varphi_0 = \pi/2$. According to the original model, this point is of special interest as the lowest-order effect of the residual Josephson energy on the pumped charge vanishes. The energy level structure at the different parameter points is given in Figs.~\ref{fig:EandV}(a--c). For the probability distribution of the ground-state population to different basis states $P(n,k)$, where $n$ refers to the island CPN state and $k$ to the feed Fock state, we only show the elements that exceed $10^{-2}$ at any time instant in Fig.~\ref{fig:EandV}(d).
At the base parameter point, there is hardly any mixing between different feed Fock subspaces which are separated by a large energy gap. Restricted to one such subspace, the populations resemble the ones given by the original model. At the second parameter point, the subspaces are no longer separated by a large gap and the energy diagram is greatly affected. Especially, the higher energy levels obtain a complicated structure having multiple avoided crossings where the energy gap is very small during the cycle. The ground-state distribution is still similar to that in the previous case, but the other adiabatic states change significantly. We give the first excited state in Fig.~\ref{fig:excited5d} showing that the state remains isolated to the first excited feed Fock subspace during the time evolution except in the vicinity of the avoided crossings where some of the population leaks to the other feed Fock subspaces. Note especially that the first excited state is no longer confined to the same Fock subspace as the ground state which may be significant for nonadiabatic and dissipative dynamics.
At the third parameter point, the parameter $g_0$ is more than 4 times that at the base point. However, the effect of this increase on the level structure seems small and the most significant changes, including the emergence of the avoided crossings at the higher levels, can be explained by the diminishing of $\Delta_f$. The ground-state distributions for the third parameter point and the base point are very similar [see Fig.~\ref{fig:EandV}(d) for the distribution at the base point], but a closer inspection reveals a small deviation between the two near the degeneracy points. This was to be expected as the terms involving the displacement operator in Eq.~(\ref{eq:H_final}) couple to the Josephson terms of the tunable junctions.

Since the robustness of the geometric charge carried by the ground state has been shown above, we turn our attention to the requirement of energy separation. Both Landau--Zener transitions\cite{pr492/1, prl107/207002, prb86/060502} and noise\cite{rmp82/1155, prl105/030401, prb84/174507, prb84/235140} may excite the system causing the transferred charge to drop. In the original model, the parameter cycle was designed such that the two lowest states were well separated from the rest of the eigenspectrum. In Fig.~\ref{fig:EandV}(a), this assumption clearly applies for our model, too. However, the assumption is broken if $\Delta_f$ is of the order of the lowest energy gap within a single Fock subspace as shown in Fig.~\ref{fig:EandV}(b). Hence, using the same driving speed as in the original model does not necessarily mean that the transitions are localized at some well-defined avoided crossings similar to the ones in Fig.~\ref{fig:EandV}(a). Furthermore, the energy separation between the higher states is very small at the emerging avoided crossings indicating that a complex process of excitations involving multiple levels may occur. This is potentially detrimental for pumping.

\section{Conclusions}  \label{sec:conclusions}

We introduced a model for the Cooper pair sluice utilized to observe the features of geometric charge transport. Our model allows the inclusion of a nonvanishing loop inductance extending the theory beyond the usually applied exact-phase-bias picture. The Hamiltonian for the device was derived by identifying relevant pairs of conjugate operators, and the current operators for the different elements of the system were formulated.

The methodology used to derive the model can be straightforwardly adapted to more complicated systems. A direct application of the methodology includes the proposal for geometric quantum computing using a loop with two islands separated by SQUIDs\cite{pra82/052304}. For this device, the practical operation window in the extended parameter space should be carefully surveyed since the ground state degeneracy might be lifted invoking error for each holonomic gate. In addition, we generally expect that using our methodology for modeling dissipative systems yields a more exact description of the effect of the environment. Applying the full quantum description is especially important for systems where the flux noise reveals important features, such as the dissipative currents flowing in the sluice\cite{prb85/024527}, or where it becomes the dominant noise source due to inbuilt robustness against charge noise, as is the case for the transmon qubit\cite{pra76/042319}.

We analyzed the charges carried by the adiabatic states during a pumping cycle both analytically and numerically. We showed that the difference in the transferred charges between our model and the exact-phase-bias picture is of the second order or higher in the strength of the coupling between the island and feed degrees of freedoms. Within experimentally feasible energy scales, we retrieve the known cosine dependence of the geometric charge with respect to the external magnetic flux with good accuracy. The geometrically transferred charge differs slightly from the original model when the additional quantum degrees of freedom are included using experimentally relevant values. The effect of altering the additional parameters included in our model on the charge depends on the external magnetic flux bias such that robust bias points emerge. The dependence differs from what the classical treatment of the loop inductance within the original model predicts and, hence, emphasizes the necessity of the quantum approach. In addition, applying the full quantum picture does not necessarily lead to a decreased geometric charge but the charge can also be increased by an appropiate selection of the loop inductance, the shunt capacitances, and the magnetic field. Even though the charge is only weakly influenced by the changes in the loop inductance, future experiments may be able to probe the variation using experimental setups based on current state-of-the-art methods\cite{prb73/214523, prl100/177201, prb86/060502}. 

Based on the geometric nature of the pumped current\cite{prb73/214523, prl100/117001}, the sluice and its generalizations have some potential to be used as quantized current sources for metrological applications. However, the metrological source requires extreme accuracy such that the maximum relative error is $\mathord{\sim} 10^{-8}$. The analysis presented in this paper shows that applying our model is necessary in the studies of the operation and feasibility of the sluice as such a source\cite{prl91/177003}. This is because experimentally relevant variations in the additional parameters included in our model, especially the self-inductance of the loop, have an effect on the pumped charge that is orders of magnitude larger than the relative error required to realize a metrological current standard.

Finally, we studied the instantaneous energy level structure and the probability distributions of the instantaneous eigenstates to different basis states near a maximally robust bias point. The energy separation of the feed Fock subspaces largely determined the instantaneous energies such that a complicated array of avoided crossings emerged when the separation was small. The ground state remained significantly robust whereas the higher states were prone to changes. Changing the additional energy scales emerging in our model was shown to be able to cause significant changes in the system energies. In particular, the energy separation between the ground state and the first excited state may be diminished and temporally altered implying that the time-local Landau-Zener transition probability is generally increased and the transitions are not necessarily localized at avoided crossings similarly to the original model. Drastic adjustment in the driving speed may be required to experimentally remain in the adiabatic limit. In addition, reduction in the energy separation between the ground state and the first excited state potentially causes the environment-induced transitions to become more prominent between these states.

\begin{acknowledgments}
The authors are grateful to S. Gasparinetti and P. Solinas for useful comments. We acknowledge the V\"ais\"al\"a Foundation, the KAUTE Foundation, and the Emil Aaltonen Foundation for financial support. This research has been supported by the Academy of Finland through its Centres of Excellence Program under Grant No. 251748 (COMP) and another Grant No. 138903 (KVANTTIYM). We have received funding from the European Research Council under Starting Independent Researcher Grant No. 278117 (SINGLEOUT).
\end{acknowledgments}

\appendix

\section{Constructing the device Hamiltonian} \label{sec:const_ham}

We denote the Cooper pair number (CPN) operators for the charge in the tunable SQUID junctions by $\hat{n}_l, \hat{n}_r$ and for the charge in the shunt capacitors by $\hat{n}_{\varphi,l}, \hat{n}_{\varphi,r}$, where the subindices $l$ and $r$ denote the left and right capacitive elements, respectively. The CPN operators for the superdonducting islands can then be written as $\hat{n} = \hat{n}_r-\hat{n}_l$, $\hat{\bar{n}}_l = \hat{n}_l - \hat{n}_{\varphi,l}$ and $\hat{\bar{n}}_r = -\hat{n}_r + \hat{n}_{\varphi,r}$ as detailed in Fig.~\ref{fig:sluice}. In the following, we refer to $\hat{n}$ as the island CPN operator and define $\hat{\bar{n}} = \frac{\hat{\bar{n}}_l-\hat{\bar{n}}_r}{2} = \frac{\hat{n}_l+\hat{n}_r}{2} - \frac{\hat{n}_{\varphi,l}+\hat{n}_{\varphi,r}}{2}$ as \emph{the feed CPN operator} describing the average charge imbalance between the superconducting islands that constitute the left and right leads. The only way to affect the imbalance is by charging the shunt capacitor system corresponding to charge transport from one capacitor to the other or by charging the capacitive junction system corresponding to charge transport between the effective junction capacitors.

The system of superconducting islands is electrically isolated so that the net charge is constant, i.e., $\hat{n} + \hat{\bar{n}}_l + \hat{\bar{n}}_r = \hat{n}_{\varphi,r} - \hat{n}_{\varphi,l} = A\hat{\mathbb{I}}_T$, where $A$ is constant in time and $\hat{\mathbb{I}}_T$ is a tensor product of the identity operators of the individual islands. The conservation reduces to a comparison of the shunt capacitors since the junction charges cancel. This means that if some charge is added to one shunt capacitor, an equivalent charge must be removed from the other. Hence, we can rewrite $\hat{\bar{n}}_r = -\hat{n}_r + \hat{n}_{\varphi,l} + A\hat{\mathbb{I}}_T$ and $\hat{\bar{n}} = \frac{\hat{\bar{n}}_l-\hat{\bar{n}}_r}{2} = \frac{\hat{n}_l+\hat{n}_r}{2} - \hat{n}_{\varphi,l} - \frac{A}{2}\hat{\mathbb{I}}_T$. We thus obtain the CPN operators for the left and right islands as $\hat{\bar{n}}_l = \hat{\bar{n}} - \frac{\hat{n}}{2} + \frac{A}{2}\hat{\mathbb{I}}_T$ and $\hat{\bar{n}}_r = -\hat{\bar{n}} - \frac{\hat{n}}{2} + \frac{A}{2}\hat{\mathbb{I}}_T$. Note that for the charge transport, the total charge in the system is irrelevant and can be taken to zero without loss of generality, that is, $\hat{n}_{\varphi,r} - \hat{n}_{\varphi,l} = A\hat{\mathbb{I}}_T = 0$. We adopt this convention. In addition, we omit explicitly writing all identity operators in the following presentation of the Hamiltonian.

Deriving the charging Hamiltonian for the device is a straigthforward task using the CPN basis formulation for the system of islands. After applying the above-mentioned transformations, the charging Hamiltonian assumes the form
\begin{equation}
\begin{split}
\hat{H}_{\mathrm{ch}} &= E_C(\hat{n}-n_g)^2 + E_{\varphi}\hat{\bar{n}}^2 + \hat{H}_{\mathrm{FIN}},
\end{split}
\end{equation}
where we have 
\begin{widetext}
\begin{equation}
\begin{split}
E_C &= 2e^2 \frac{(2C_{\varphi}+C_g)(2C_{\varphi}+C_l+C_r)}{4\{ C_gC_lC_r + C_{\varphi}^2(C_g + C_l + C_r) + C_{\varphi}[2C_lC_r + C_g(C_l+C_r)]\}}, \\
E_{\varphi} &= 2e^2 \frac{C_g(C_l+C_r)+2C_{\varphi}(C_g+C_l+C_r)}{C_gC_lC_r + C_{\varphi}^2(C_g + C_l + C_r) + C_{\varphi}[2C_lC_r + C_g(C_l+C_r)]}.
\end{split}
\label{eq:charging_normal}
\end{equation}
\end{widetext}
Here $C_g$ is the gate capacitance, $C_{\varphi}$ is the capacitance of one of the identical shunt capacitors and $C_l,C_r$ are the internal capacitances of the tunable junctions. Furthermore, the normalized gate charge is $n_g = V_gC_g/(2e)$, where $V_g$ is the externally controlled gate voltage. The final term in the Hamiltonian is
\begin{equation}
\begin{split}
\hat{H}_{\mathrm{FIN}} = E_1 n_g\hat{n} + E_2 n_g^2 + E_3 n_g\hat{\bar{n}}  + E_4\hat{n}\hat{\bar{n}},
\end{split}
\label{eq:HFIN}
\end{equation}
where the prefactors are 
\begin{widetext}
\begin{equation}
\begin{split}
E_1 &= 2e^2 \frac{C_g(2C_{\varphi}+C_l+C_r)}{2\{ C_gC_lC_r + C_{\varphi}^2(C_g + C_l + C_r) + C_{\varphi}[2C_lC_r + C_g(C_l+C_r)]\} }, \\
E_2 &= 2e^2 \frac{4C_lC_r - C_g(C_l+C_r) + 2C_{\varphi}(C_l + C_r - C_g)}{4\{ C_gC_lC_r + C_{\varphi}^2(C_g + C_l + C_r) + C_{\varphi}[2C_lC_r + C_g(C_l+C_r)]\} }, \\
E_3 &= 2e^2 \frac{2C_{\varphi}(C_r-C_l)}{C_gC_lC_r + C_{\varphi}^2(C_g + C_l + C_r) + C_{\varphi}[2C_lC_r + C_g(C_l+C_r)]}, \\
E_4 &= 2e^2 \frac{(2C_{\varphi}+C_g)(C_l-C_r)}{C_gC_lC_r + C_{\varphi}^2(C_g + C_l + C_r) + C_{\varphi}[2C_lC_r + C_g(C_l+C_r)]}.
\end{split}
\label{eq:charging_HFIN}
\end{equation}
\end{widetext}
Presenting the Hamiltonian in this form showcases its properties. If the island is capacitively symmetric $C_l = C_r = C_J$, we have $E_3 = E_4 = 0$ so that the first-order feed term and the term coupling the island and feed degrees of freedoms vanish from the charging Hamiltonian. Furthermore, the remaining prefactors become $E_C = 2e^2 \frac{(2C_{\varphi}+C_g)(C_{\varphi}+C_J)}{2[C_gC_J^2 + C_{\varphi}^2(C_g + 2C_J) + 2C_{\varphi}C_J(C_J + C_g)]}$, $E_{\varphi} = 2e^2 \frac{2(C_gC_J + C_{\varphi}(C_g+2C_J))}{C_gC_J^2 + C_{\varphi}^2(C_g + 2C_J) + 2C_{\varphi}C_J(C_J + C_g)}$, $E_1 = 2e^2 \frac{C_g(C_{\varphi}+C_J)}{C_gC_J^2 + C_{\varphi}^2(C_g + 2C_J) + 2C_{\varphi}C_J(C_J + C_g)}$ and $E_2 = 2e^2 \frac{2C_J^2 - C_gC_J + C_{\varphi}(2C_J - C_g)}{2[C_gC_J^2 + C_{\varphi}^2(C_g + 2C_J) + 2C_{\varphi}C_J(C_J + C_g)]}$. If we assume that the gate capacitance is vanishingly small compared to the shunt capacitance, the first order island charge term becomes negligible. The remaining term $E_2n_g^2$ in $\hat{H}_{\mathrm{FIN}}$ is proportional to the identity operator, and hence only induces a time-dependent shift of the zero point of the energy that can be neglected.

The inductive part of the Hamiltonian stems from the inductive energy $\hat{H}_{\mathrm{ind}} = \frac{1}{2} L\hat{I}_L^2$ where $L$ is the total inductance of the loop element and $\hat{I}_L$ is the operator for the current passing through it. Note that in addition to the geometric inductance $L_G$, $L = L_G + L_K$ also includes the kinetic inductance $L_K$ which can potentially be significant for a superconducting wire\cite{nanotech21/445202}. The total magnetic flux threading the loop can be expressed as a sum of the applied external flux and the flux induced by the supercurrent due to the geometric inductance $\hat{\Phi} = \Phi_{\mathrm{ext}} + L_G\hat{I}_L$. Furthermore, the operator for the total phase difference over all the weak links in the loop $\hat{\varphi}$ is related to the total flux operator by $\frac{2\pi \hat{\Phi}}{\Phi_0} = \hat{\varphi}_l + \hat{\varphi}_r - \frac{2e}{\hbar}L_K\hat{I}_L = \hat{\varphi} - \frac{2e}{\hbar}L_K\hat{I}_L$, where $\hat{\varphi}_l$ and $\hat{\varphi}_r$ are the operators for the gauge-invariant phase differences over the left and right SQUIDs, respectively. The kinetic inductance adds to the total phase difference over the loop according to the ac Josephson equation\cite{ItS}, and the directions over which the phase differences and the inductive current are calculated are given in Figs.~\ref{fig:sluice}~and~\ref{fig:current_schematic}. Using the above-mentioned identities, the operator for the inductive current assumes the form
\begin{equation}
\begin{split}
\hat{I}_L = \frac{\Phi_0}{2\pi} \frac{1}{L} (\hat{\varphi}-\varphi_0),
\end{split}
\label{eq:IL_basic}
\end{equation}
where $\varphi_{0} = \frac{2\pi \Phi_{\mathrm{ext}}}{\Phi_0}$ and we denote the flux quantum by $\Phi_0 = h/(2e)$. Thus, the inductive Hamiltonian can be written as
\begin{equation}
\begin{split}
\hat{H}_{\mathrm{ind}} = \frac{E_L}{2}(\hat{\varphi}-\varphi_0)^2,
\end{split}
\end{equation}
where we denote $E_L = \frac{1}{L} \left( \frac{\Phi_0}{2\pi} \right)^2$.

Finally, the Josephson part describing the tunneling through the junctions is given by\cite{ItS}
\begin{equation}
\begin{split}
\hat{H}_{J} = -E_{Jr} \cos \hat{\varphi}_r - E_{Jl} \cos \hat{\varphi}_l,
\end{split}
\end{equation}
where $E_{Jl}(\Phi_l)$ and $E_{Jr}(\Phi_r)$ are the Josephson energies for the left and right tunable junctions, respectively. Starting from the operators for the phase differences over the SQUIDs, we perform a transformation to a new pair of phase operators such that $\hat{\varphi} = \hat{\varphi}_l + \hat{\varphi}_r$ and $\hat{\phi} = (\hat{\varphi}_r - \hat{\varphi}_l)/2$. This transformation is well-founded since the canonical commutation relation between the charge and phase operators of the individual junctions yields $[\hat{\phi},\hat{n}] = i$ indicating that $\hat{\phi}$ is the island phase operator and $[\hat{\varphi},\hat{\bar{n}}] = i$ indicating that $\hat{\varphi}$ is the canonical conjugate of the feed CPN operator\footnote{We can write $e^{i\hat{\varphi}_l} = \sum_{n,\bar{n}_l} \ket{n-1,\bar{n}_l+1}\bra{\bar{n}_l,n}$ and $e^{i\hat{\varphi}_r} = \sum_{n,\bar{n}_r} \ket{n+1,\bar{n}_r-1}\bra{\bar{n}_r,n}$. The commutation of the individual phases yields $e^{i\hat{\varphi}} = \sum_{n,\bar{n}_l,\bar{n}_r} \ket{n,\bar{n}_l+1,\bar{n}_r-1}\bra{\bar{n}_r,\bar{n}_l,n}$ implying that $e^{i\hat{\varphi}} = \sum_{n,\bar{n}} \ket{n,\bar{n}+1}\bra{\bar{n},n}$. Thus, $[\hat{\varphi},\hat{\bar{n}}] = i$.}. If the average charge imbalance between the left and right leads changes by one, the occupation in the eigenbasis of $\hat{\bar{n}}$ subsequently changes. This implies charge transfer in the system and allows us to monitor the system currents. With the transformed phases, the Josephson part assumes the form
\begin{equation}
\begin{split}
\hat{H}_{J} = -E_{Jr} \cos \left( \frac{\hat{\varphi}}{2} + \hat{\phi} \right)  - E_{Jl} \cos \left( \frac{\hat{\varphi}}{2} - \hat{\phi} \right).
\end{split}
\end{equation}

\section{Perturbative analysis} \label{sec:firstorder}

The unperturbed states $\ket{k}\ket{\Psi_n}$ are defined via the eigenproblem $[\hat{H}_S \otimes \hat{\mathbb{I}}_{\varphi} + \hat{\mathbb{I}}_n \otimes \hat{H}_{\varphi}]\ket{k}\ket{\Psi_n} = [\epsilon_k + \sqrt{2E_{\varphi}E_L} \left( n+\frac{1}{2} \right)]\ket{k}\ket{\Psi_n}$ where $\ket{k}$ is an eigenstate of $\hat{H}_S$ and $\epsilon_k$ is the corresponding eigenenergy. Denoting the adiabatic states and energies as $\ket{\Phi_{kn}}$ and $E_{kn}$ corresponding to the $kn$th unperturbed state, the inner products relating to the different current elements mentioned in Sec.~\ref{subsec:perturbative} become
\begin{widetext}
\begin{equation}
\begin{split}
\braket{\Phi_{kn}|\hat{I}_I^{(0)}|\Phi_{pm}} &= \braket{k|\hat{I}_S|p} \delta_{n,m} + g_0 \frac{\hbar}{e} \sum_l \braket{k|\hat{I}_S|l} \braket{l|\hat{I}_S|p} \left[ \left( \frac{\sqrt{m}}{\epsilon_p-\epsilon_l+\sqrt{2E_{\varphi}E_L}} + \frac{\sqrt{m}}{\epsilon_k-\epsilon_l-\sqrt{2E_{\varphi}E_L}} \right) \delta_{n,m-1} \right. \\ &+ \left. \left( \frac{\sqrt{m+1}}{\epsilon_p-\epsilon_l-\sqrt{2E_{\varphi}E_L}} + \frac{\sqrt{m+1}}{\epsilon_k-\epsilon_l+\sqrt{2E_{\varphi}E_L}} \right) \delta_{n,m+1} \right] + O(g_0^2),
\end{split}
\label{eq:I_I^0}
\end{equation}
\end{widetext}
and
\begin{widetext}
\begin{equation}
\begin{split}
\braket{\Phi_{kn}|\hat{I}_I^{(1)}|\Phi_{pm}} &= - g_0 \frac{e}{\hbar} \braket{k|[\hat{H}_B + \hat{H}_B^{\dagger}]|p} (\sqrt{m}\delta_{n,m-1} + \sqrt{m+1}\delta_{n,m+1}) + O(g_0^2),
\end{split}
\label{eq:I_I^1}
\end{equation}
\end{widetext}
and
\begin{widetext}
\begin{equation}
\begin{split}
\braket{\Phi_{kn}|\hat{I}_L|\Phi_{pm}} = \braket{k|\hat{I}_S|p} \frac{2E_{\varphi}E_L}{(\epsilon_p-\epsilon_k)^2 - 2E_{\varphi}E_L} \delta_{n,m} + \frac{e}{\hbar}(8E_{\varphi}E_L^3)^{\frac{1}{4}}\delta_{k,p}(\sqrt{m}\delta_{n,m-1} + \sqrt{m+1}\delta_{n,m+1}) + O(g_0^2).
\end{split}
\label{eq:I_L_B}
\end{equation}
\end{widetext}
Using Eqs.~(\ref{eq:I_I^0}) and (\ref{eq:I_I^1}), the dynamical current through the island carried by $\ket{\Phi_{kn}}$ can be readily written as $I_I^{(D,kn)} = \braket{\Phi_{kn}|\hat{I}_I|\Phi_{kn}} = \braket{k|\hat{I}_S|k} + O(g_0^2)$. Similarly, Eq.~(\ref{eq:I_L_B}) allows us to write $I_L^{(D,kn)} = \braket{\Phi_{kn}|\hat{I}_L|\Phi_{kn}} = -\braket{k|\hat{I}_S|k} + O(g_0^2)$.

In order to study the geometric currents, we need the temporal derivatives of the adiabatic states. The corresponding inner product is
\begin{widetext}
\begin{equation}
\begin{split}
\braket{\Phi_{pm}|\partial_t|\Phi_{kn}} &= \braket{p|\dot{k}}\delta_{n,m} + g_0\frac{\hbar}{e} \sum_l \left[ \braket{p|\dot{l}}\braket{l|\hat{I}_S|k} \left( \frac{\sqrt{m+1}\delta_{n,m+1}}{\epsilon_k-\epsilon_l+\sqrt{2E_{\varphi}E_L}} + \frac{\sqrt{m}\delta_{n,m-1}}{\epsilon_k-\epsilon_l-\sqrt{2E_{\varphi}E_L}} \right) \right. \\ &+ \left. \braket{l|\dot{k}}\braket{p|\hat{I}_S|l} \left( \frac{\sqrt{m}\delta_{n,m-1}}{\epsilon_p-\epsilon_l+\sqrt{2E_{\varphi}E_L}} + \frac{\sqrt{m+1}\delta_{n,m+1}}{\epsilon_p-\epsilon_l-\sqrt{2E_{\varphi}E_L}} \right) \right] \\ &+ g_0\frac{\hbar}{e} \partial_t \left( \frac{\braket{p|\hat{I}_S|k}}{\epsilon_k-\epsilon_p+\sqrt{2E_{\varphi}E_L}} \right) \sqrt{m+1}\delta_{n,m+1} + g_0\frac{\hbar}{e} \partial_t \left( \frac{\braket{p|\hat{I}_S|k}}{\epsilon_k-\epsilon_p-\sqrt{2E_{\varphi}E_L}} \right) \sqrt{m}\delta_{n,m-1} \\ &+ O(g_0^2),
\end{split}
\label{eq:det_inn}
\end{equation}
\end{widetext}
where the dot indicates time derivative. Using Eq.~(\ref{eq:det_inn}) and the above-mentioned current elements, the geometric current through the island becomes 
\begin{widetext}
\begin{equation}
\begin{split}
I_I^{(G,kn)} = 2\hbar \Im m \sum_{p,m}^* \frac{\braket{\Phi_{kn}|\hat{I}_I|\Phi_{pm}}}{E_{kn}-E_{pm}} \braket{\Phi_{pm}|\partial_t|\Phi_{kn}} = 2\hbar \Im m \sum_{p\neq k} \frac{\braket{k|\hat{I}_S|p}}{\epsilon_k-\epsilon_p}\braket{p|\dot{k}} + O(g_0^2),
\end{split}
\label{eq:I_I^G}
\end{equation}
\end{widetext}
where $\sum^*$ denotes a double summation that does not include the term for which $p=k$ and $m=n$ simultaneously. Calculating the inductive element in a similar manner, we obtain $I_L^{(G,kn)} = - I_I^{(G,kn)}$ up to the first order.

\section{Parameter cycle} \label{sec:cycle}

The parameter cycle consists of six equivalently long legs designed to maintain good charge quantization per cycle. In the beginning, both SQUIDs are closed so that their residual Josephson energies are $E_{Jl}^{\mathrm{min}}$ and $E_{Jr}^{\mathrm{min}}$, and the gate charge is $n_g^{\mathrm{min}}$. Nonzero values for the residual Josephson energies are typically used to include the effect of nonidealities stemming from the manufacturing of the junctions, and lead to a nonvanishing supercurrent flowing through the device. During the first stage, the left SQUID is opened by bringing its Josephson energy linearly to $E_{Jl}^{\mathrm{max}}$. During the second stage, the gate charge is linerly increased to $n_g^{\mathrm{max}}$ drawing charge from the left lead to the centre island. The left SQUID is then linearly closed and the right SQUID opened in a similar manner bringing its Josephson energy to $E_{Jr}^{\mathrm{max}}$. During the fifth stage, the gate charge is brought linearly back to $n_g^{\mathrm{min}}$ pushing the charge to the right lead. Finally, the right SQUID is linearly closed and the system has completed a closed loop in the external parameter space.

\bibliography{localbib.bib}

\end{document}